\documentclass[aps,prb,onecolumn,showpacs,superscriptaddress,groupedaddress,longbibliography]{revtex4-1}
\usepackage[dvips]{graphics,graphicx}
\usepackage{subfig}
\usepackage{bbold}
\usepackage{amsmath}
\usepackage{amsfonts}
\usepackage{bm}
\usepackage{mathrsfs}
\usepackage{epstopdf}
\usepackage{hyperref}
\usepackage{amsmath,amsfonts,amssymb,amsthm}
\newcommand{\scs}{\scriptscriptstyle}

\begin{document}

\title{Optical bistability with bound states in the continuum in dielectric gratings}

\author{Dmitrii N. Maksimov$^{1,2}$}
\author{Andrey A. Bogdanov$^3$}
\author{Evgeny N. Bulgakov$^{1,4}$}

\affiliation{$^1$Kirensky Institute of Physics, Federal Research Center KSC SB RAS, 660036, Krasnoyarsk, Russia\\
$^2$Siberian Federal University, 660041, Krasnoyarsk, Russia\\
$^3$Department of Physics and Engineering, ITMO University, 191002, St. Petersburg, Russia\\
$^4$Reshetnev Siberian State University of Science and Technology, 660037, Krasnoyarsk, Russia}

\date{\today}
\begin{abstract}
We consider light scattering by dielectric gratings supporting optical bound states in the continuum. Due to the presence of instantaneous
Kerr nonlinearity  the critical field enhancement in the spectral vicinity of the bound state triggers the effect of optical bistability.
The onset of bistability is explained theoretically in the framework
of the temporal coupled mode theory. As the central result
we cast the problem into the form of a singly field-driven nonlinear oscillator. The theoretical results are verified in comparison against
full-wave numerical simulations.
\end{abstract}
 \maketitle

\section{Introduction}

Engineering high{-}quality resonances which provide access to tightly localized optical fields has become a topic of
paramount importance in electromagnetism \cite{John12, Marpaung13, Qiao18, Hsu16}. In that context{,} dielectric gratings
(DGs) are a useful optical instrument with numerous applications {relying}
on high{-}quality resonances \cite{Chang-Hasnain12, Qiao18} that occur in the spectral vicinity of the avoided
crossings of the DG modes \cite{Karagodsky11}. The utmost case of light localization is a bound state in the continuum (BIC) -
an embedded state with infinite quality factor coexisting with the scattering solutions \cite{Hsu16, Koshelev19}. Since the seminal
paper by Marinica, Borisov, and Shabanov \cite{Marinica08} BICs in all-dielectric DGs have been extensively studied
 both theoretically \cite{ Monticone17, Bulgakov18, Bulgakov18a,  Lee19, Bykov19, Gao19} and experimentally  \cite{Sadrieva17, Hemmati19}.
  Lately, optical BICs have also been reported in hybrid photonic-plasmonic gratings \cite{Azzam18, Kikkawa19}.

The BICs are spectrally surrounded by a leaky band of high{-}quality resonances which can be excited from the far-zone
\cite{Yuan17}. The excitation of the strong resonances leads to critical field enhancement
\cite{Yoon15, Mocella15a} with the near-field amplitude controlled by the frequency and the angle of incidence of the incoming monochromatic wave.
The critical field enhancement allows for activating nonlinear optical effects even with a small amplitude of the incident waves.
Such resonant enhancement of nonlinear effects may lead to the effects of symmetry breaking \cite{Bulgakov11} and channel dropping \cite{Bulgakov13}.

Among various potential applications in nonlinear optics the BICs have been used
for second harmonic (SH) generation.
In particular, giant conversion efficiency into SH (up to 40\%) was predicted for an array of parallel dielectric
cylinders~\cite{ndangali2013resonant}. A more practical design of AlGaAs metasurface on a quartz  substrate supporting BIC was analyzed in
\cite{koshelev2019meta}, where the efficiency of SH generation $P_{2\omega}/P_{\omega}^2\sim10^{-2}$~W is predicted in the vicinity of a BIC.
 Recently, it was shown theoretically that BIC can enhance the SH convention efficiency in transition-metal dichalcogenide monolayers
 by more than four orders of magnitude~\cite{wang2018large}. The BIC is a dark (optically inactive) resonance which cannot be excited from
 the far field, however, it was shown in \cite{yuan2020excitation} that BICs in periodic dielectric structures can be excited by non-linear
  polarization at the SH frequency induced by the incident field. The same mechanism of destructive interference underlying BICs can result
  in appearance of high-quality modes in subwavelength dielectric resonators~\cite{rybin2017high, bogdanov2019bound}, which also
  demonstrate giant SH generation efficiency~\cite{carletti2018giant, koshelev2020subwavelength}.

In this paper we consider the effect of the critical field
enhancement on optical bistability induced by instantaneous Kerr nonlinearity. Such optical bistability emerges in
the scattering spectra
in the form of nonlinear Fano resonances \cite{Miroshnichenko05, Shipman12}. Previously the studies of optical bistability with BICs
solely relied on either brute force full-wave modelling \cite{Yuan16, Yuan17} or phenomenological
coupled{-}mode approach \cite{Krasikov18}. Recently, having considered
{an} array of nonlinear cylinders, we combined the
two approaches into a single theory \cite{Bulgakov19} that reduces
the problem of finding the nonlinear response to solving a
nonlinear coupled{-}mode equation for a single variable.
Herewith all the parameters of the coupled{-}mode equation are known
from solving the linear scattering problem in the spectral vicinity of
the BIC which is a far easier task than full-wave modelling of nonlinear Maxwell's equation. In this paper{,} we present a generic
theory applicable to planar structure with no mirror symmetry with
respect to reflection in the plane of the structure. The theory
is verified in comparison against full-wave numerical solutions
of Maxwell's equations.

\begin{figure*}[t]
\centering\includegraphics[width=1\textwidth, height=0.4\textwidth, trim={4cm 3cm 3cm 3cm},clip]{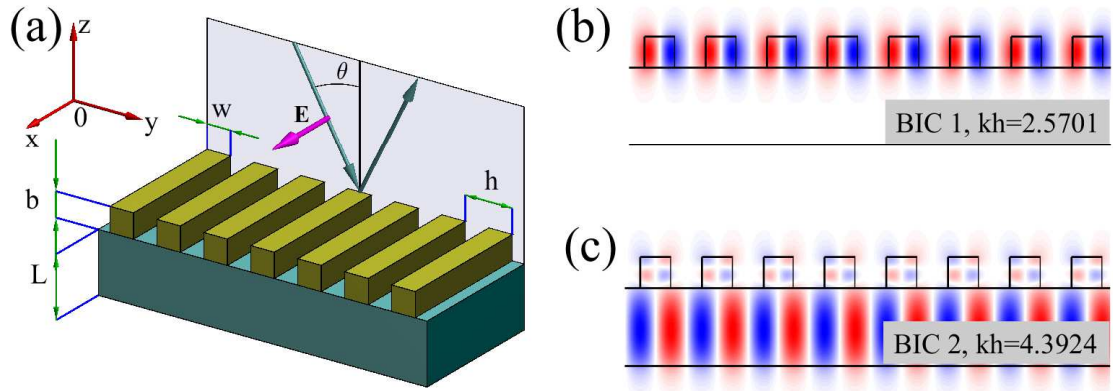}
\caption{(a) The dielectric grating of Si bars on glass substrate. The
plane of incidence $y0z$ is shaded grey. The magenta arrow shows the
electric vector of the incident wave. The parameters are $w=0.5h$, $b=0.5h$, $L=1.25h$ (b, c) The electric field profiles of two symmetry protected BICs visualized as $E_x$ in the $y0z$ plane.} \label{fig1}
\end{figure*}

\section{Bound states in the continuum}
The system under consideration is shown in Fig.~\ref{fig1}~(a). It is a dielectric grating assembled of rectangular dielectric bars made of Si. The bars are periodically placed on the
glass substrate.  Here we only consider the scattering of TE polarized waves with the electric vector aligned with the Si bars as shown in Fig. \ref{fig1}. Under such conditions
the propagation of electromagnetic waves is controlled by the Helmholtz equation for
the $x$ component of the electric field
\begin{equation}\label{Helmholtz}
\left(\frac{\partial^2}{\partial y^2}+ \frac{\partial^2}{\partial z^2}\right)E_x+k^2[n_0^2+2n_0n_2|E_x|^2]E_x=0,
\end{equation}
where $k$ is the vacuum wavenumber, and $n_{0,2}$ are linear and nonlinear refractive indices, correspondingly. In what follows the refractive index of Si is taken as $n_0=3.575$, while the refractive index of the substrate $n_0=1.5$.

Our numerical simulations with the use of Dirichlet-to-Neumann map method \cite{Huang06} have demonstrated that for the set of parameters specified in the caption to Fig. \ref{fig1} the system supports two in-$\Gamma$ BICs coexisting with the zeroth diffraction order. The eigenmode profiles of the BICs are shown in Fig.~\ref{fig1}~(b) and Fig.~\ref{fig1}~(c). Notice the striking difference between BIC 1 and BIC 2 in Fig.~\ref{fig1}~(b) and Fig.~\ref{fig1}~(c): the field of BIC 1 is mostly localized in the Si bars, whereas the field of BIC 2 is spread across the whole grating. This difference is due to the higher
eigenfrequency of BIC 2 allowing the first diffraction order in the glass substrate.

\begin{figure*}[t]
\centering\includegraphics[width=1\textwidth, height=0.4\textwidth, trim={4cm 1cm 5cm 2cm},clip]{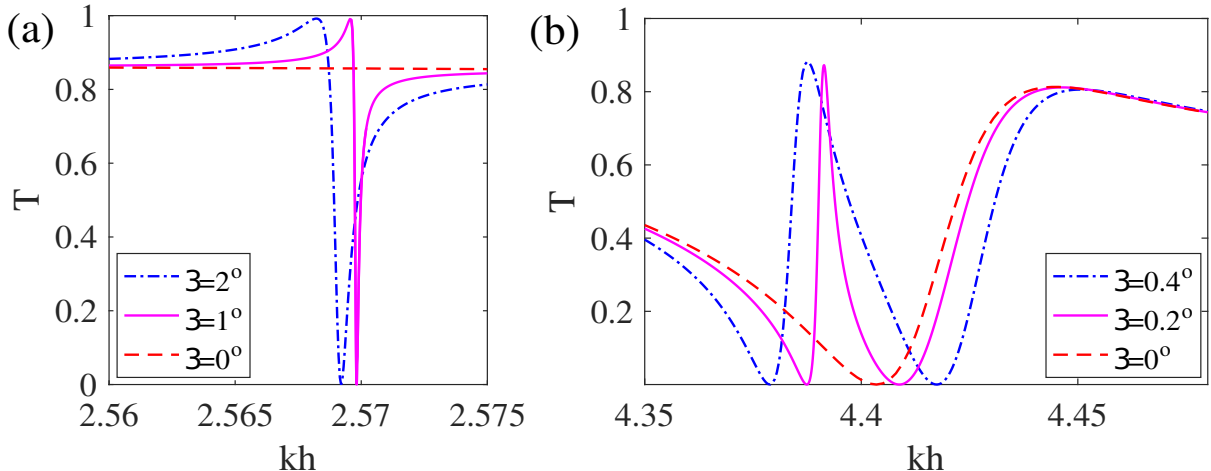}
\caption{(a) Transmittance spectrum in vicinity of BIC 1 at three different angles of incidence specified in the inset. (b) Transmittance spectrum in vicinity of BIC 2 at three different angles of incidence specified in the inset.}\label{fig2}
\end{figure*}

One important property of the BICs is the emergence of a collapsing Fano feature in its parametric vicinity \cite{Kim,Shipman,SBR,Blanchard16,Bulgakov18b}. In this paper, we consider light
scattering near the normal incidence so that the incident light couples
to the band of resonant modes with diverging $Q$-factor as $\theta\rightarrow 0$. In Fig.~\ref{fig2}~(a) and Fig.~\ref{fig2}~(b) we show the transmittance spectra in the spectral vicinity of BIC 1 and BIC 2, correspondingly. One can see that in both cases one observes a narrow
Fano feature which collapses at the exact normal incidence. There is also a difference between the two cases. Namely, there is more than a single
Fano resonance at BIC 2. The transmittance exhibits two zeros and an extra peak which does not vanish at the normal incidence. This difference can be explained through different nature of BIC 1 and BIC 2 One can see from Fig. \ref{fig2} that BIC 1 occurs as an isolated resonance, while BIC 2 emerges as a result of hybridisation of two resonant modes with one of them acquiring infinite life-time.
The latter mechanism of BIC has been previously demonstrated for dielectric gratings in \cite{Bulgakov18a}. In the next section we provide a theoretical description of the lineshapes of the Fano anomalies induced by the BICs extended to the effects of Kerr nonlinearity triggered by critical field enhancement.

\section{Scattering theory}
The aim of this section is to formulate the equation for
the amplitude of the quasi-BIC resonant mode in the framework of
the temporal coupled mode theory (TCMT) \cite{Fan03}.
The generic case of a TCMT applied to a 2D structure is considered in \cite{Zhou16}. It has been demonstrated that in the absence of
mirror symmetry with respect to $y\rightarrow-y$ the application of TCMT requires considering four scattering channels. However, as the above symmetry holds in our case, we shall apply a two-channel TCMT in this paper.
Following \cite{Bulgakov19} we only consider the effect of a single resonant mode
mentioning in passing that a generalization of the TCMT to multimodal case is possible \cite{Suh04}.

\subsection{Coupled mode approach}
Let us start with the TCMT equation for an isolated resonance \cite{Fan03}
\begin{equation}\label{CMT}
    \frac{da(t)}{dt}=i(\omega_0-\Gamma)a+\langle d^*| s^{\scs(+)}\rangle,
\end{equation}
where $a(t)$ is the time-dependent amplitude of the resonant mode, $\omega_0$ is the resonant frequency, $\Gamma$ is the inverse life-time of the resonance, $|d\rangle$ is the vectors of coupling constants to
the scattering channels, and $| s^{\scs(+)}\rangle$ is the vector of
incident amplitudes. Since we stay in the domain where only specular reflection is allowed, both  $| s^{\scs(+)}\rangle=(s^{\scs(+)}_1, s^{\scs(+)}_2)^{\intercal}$, and
$|d\rangle=\left(d_1, d_2\right)^{\intercal}$ are $2\times1$ vectors. The subscripts $1,2$ are applied to the upper and lower half-spaces, correspondingly. Let us, e.g., assume that a monochromatic plane wave with frequency $\omega$ is incident onto the grating from the upper half-space.
The vector of the incident amplitudes is written as
\begin{equation}
    | s^{\scs(+)}\rangle=\left(\sqrt{I_0}, 0\right)^{\intercal},
\end{equation}
where ${I_0}$ is the flux density supported by the incident wave.
After
the time-harmonic substitution $a(t)=ae^{i\omega t}$ one finds
\begin{equation} \label{amplitude}
    a=\frac{d_1\sqrt{I_0}}{[i(\omega-\omega_0)+\Gamma]}.
\end{equation}
Finally, the outgoing amplitudes can be found from the following equation
\begin{equation}
    | s^{\scs (-)}\rangle=\widehat{C}+a|d\rangle.
\end{equation}
Here, $\widehat{C}$ is the matrix of direct (non-resonant) process.
In the case of the symmetry protected BIC, the matrix $\widehat{C}$ can be easily obtained numerically by solving the scattering problem at the normal incidence with the incident frequency equal to the BIC frequency. In the other words exactly in the point
of the Fano resonance collapse \cite{Bulgakov19}.

The general solution of the linear scattering problem can be written through the scattering matrix $\widehat{S}(\omega)$  which links the vectors of incident and outgoing amplitudes
\begin{equation}\label{out}
| s^{\scs (-)}\rangle=\widehat{S}({\omega})| s^{\scs (+)}\rangle.
\end{equation}
Since the system under consideration is both energy preserving and symmetric with respect to time reversal, the matrices $\widehat{S}(\omega)$ and $\widehat{C}$ are simultaneously unitary and symmetric \cite{Zhao19}.
The most generic form of $\widehat{C}$ can be parameterized in the following manner
\begin{equation}\label{dir}
    \widehat{C}=e^{i\phi}
    \left(
    \begin{array}{cc}
        \rho e^{-i\eta} & i\tau \\
           i\tau  &  \rho e^{i\eta}
    \end{array}
    \right),
\end{equation}
where the real valued $\rho$ and $\tau$ are the absolute values of
the reflection and transmission amplitudes which have to satisfy the following
equation
\begin{equation}
   \rho^2+\tau^2=1.
\end{equation}
Thus, taking the above into account we are left with only
three independent parameters, $\theta, \eta$, and $\rho$ which
can be analytically derived from Eq. (\ref{dir}).

The quantities $\widehat{C}$ and $|d\rangle$ are linked
through the following equation \cite{Fan03}
\begin{equation}\label{cond2}
    \widehat{C}|d^*\rangle=-|d\rangle,
\end{equation}
which is a consequence of both energy conservation and time-reversal symmetry.
Equation (\ref{cond2}) constitutes a homogeneous algebraic equation for unknown $|d\rangle$. Since the complex conjugation is involved in Eq. (\ref{cond2}) it has to be solved for four independent variables, i.e. the real and imaginary parts of $|d\rangle$. This results in a system
of four equation of rank $2$. Therefore
The general solution of Eq. (\ref{cond2})
can be written as a function of two independent parameters $\alpha$ and $\beta$.
\begin{equation}\label{d}
|d\rangle=
\left(
\begin{array}{c}
     (\tau \alpha- i(1+\rho)\beta)e^{i\frac{\phi-\eta}{2}}\\
     (\tau \beta- i(1+\rho)\alpha)e^{i\frac{\phi+\eta}{2}}
\end{array}
\right).
\end{equation}
Notice that in general $d_1 \neq d_2$. Thus, Eq. (\ref{d}) takes into account the asymmetry of the coupling to the upper and lower half-spaces due to the lack of mirror symmetry in the plane of the structure, see Fig.~\ref{fig1}~(a).

Another important relationship \cite{Fan03} is a also a consequence of energy conservation
\begin{equation}\label{cond3}
    2\Gamma=\langle d|d\rangle.
\end{equation}
The above equation is derived by considering the decay dynamics of the
system with no impinging wave. Assume that a certain amount of energy $E$
is load into the resonant mode, then the solution of Eq. (\ref{CMT})
$a(t)=a(0)e^{i\omega_0t-\Gamma t}$. Given that the eigenmode stores a unit energy, the energy dissipation rate can be found as
\begin{equation}\label{rate1}
    \frac{d {\cal E}}{dt}=-2\Gamma |a_0|^2,
\end{equation}
where ${\cal E}$ is the energy stored in the resonant eigenmode. On the other hand if each scattering channel attenuates a unit of energy per unit of time, Eq. (\ref{out}) yields
\begin{equation}\label{rate2}
    \frac{d {\cal E}}{dt}= -\langle d|d\rangle|a_0|^2.
\end{equation}
Combining Eqs. (\ref{rate1}) and (\ref{rate2}) we find Eq. (\ref{cond3}). Notice, that normalization of both the eigenmode
and decay channels is important for deriving Eq.~(\ref{cond3}).
Application of Eq.~(\ref{cond3}) to Eq.~(\ref{d})
yields
\begin{equation}\label{ab}
\alpha^2+\beta^2=\frac{2\Gamma}{\tau^2+(1+\rho)^2}.
\end{equation}

Let us summarize the findings of this subsection. First,
as it is seen from Eq. (\ref{amplitude}) the resonant response
is due to vanishing $\Gamma$ in denominator. Notice that both $\Gamma$ and $\omega_0$ are known from the eigenmode spectrum
of the grating, they can be determined as the real and imaginary part of the resonant frequency of the leaky band
host the BIC, as it has been done in \cite{Bulgakov19}. Second,
the coupling vector $|d\rangle$ is defined from the matrix of the direct process Eq. (\ref{dir}) via Eq. (\ref{d}) up to two unknown real-valued parameters. Quite remarkable is that
the presence of a symmetry protected BIC gives an easy access to the the matrix
of the direct process by simply computing the scattering solution at the BIC frequency and the normal incidence \cite{Bulgakov19} . Finally, Eq. (\ref{ab}) allows for eliminating of one of the free parameters, say $\alpha$ in Eq. (\ref{d}). The remaining parameter $\beta$ can be found by easily found by fitting the
transmittance spectrum found through the full-wave solution
of the scattering problem.

\subsection{Green's function}
Let us now generalize the above result to the system with Kerr nonlinearity.  In this subsection we apply the resonant state expansion method \cite{Weiss17} for deriving the TCMT equation with account of nonlinearity. The key figure of merit in the resonant state expansion method is Green's function of Maxwell's equations.
According to \cite{Weiss17} the spectral representation of Green's function can be written as
\begin{equation}\label{Green}
G({\bf r}', {\bf r}, k_0, k_y)=
\int\kern-1.5em\sum_n
\frac{E_x^{(n)}({\bf r}, k_y)E_x^{(n)}({\bf r'}, -k_y)}
{2k[k-k_n(k_y)]},
\end{equation}
where $E_x^{(n)}({\bf r}, k_y)$ the field profile of
the $n_{\rm th}$ resonant eigenmode and $k_n(k_y)$ is
the dispersion of the resonant eigenfrequency of the leaky band in terms of vacuum wave number, $k=\omega/c$ with $c$ as the speed of light. The symbol $\int\kern-1em\sum_n$ is used for the combined contribution of a discrete sum and integration along the cuts. For the spectral representation Eq. (\ref{Green})
to be valid the eigenfields $E_x^{(n)}({\bf r}, k_y)$
must obey the following normalization condition
\begin{equation}\label{norm}
1+\delta_{0,k_n}=I_n^{V}+\lim_{k\rightarrow k_n}\frac{S_n^{\scs \partial V}}{k^2-k_n^2}
\end{equation}
with
\begin{equation}\label{volume}
I_n^{V}=\int_VdV{E_x^{(n)}({\bf r},-k_y)E_x^{(n)}({\bf r'}, k_y)}
\end{equation}
and
\begin{equation}\label{surface}
S_n^{\scs \partial V}=\oint_{\partial V}dS\left[
E_x^{(n)}({\bf r},-k_y)\partial_S\tilde{E}_x^{(n)}({\bf r'}, k_y,k)
-\tilde{E}_x^{(n)}({\bf r},-k_y,k)\partial_S{E}_x^{(n)}({\bf r'}, k_y)\right],
\end{equation}
where $\partial_S$ is used for the normal derivative with respect to the boundary of the
elementary cell and $\tilde{E}_z^{(n)}({\bf r'}, k_y,k_0)$ is the analytic continuation
of the eigenfield in the vicinity of its resonant eigenfrequency such as
\begin{equation}\label{continuation}
 {E}_x^{(n)}({\bf r'}, k_y)=\lim_{k\rightarrow k_n} \tilde{E}_x^{(n)}({\bf r'}, k_y,k).
\end{equation}

\subsection{Resonant approximation}
To establish a link between the resonant state expansion and the single mode TCMT we apply resonant approximation, i.e. in Eq. (\ref{Green}) we retain only the term with $k-k_n(k_y)$ in the denominator. All the other terms are assumed to be independent
of frequency on the scale of the narrow Fano feature induced by the BIC. In terms of the TCMT the non-resonant terms are accumulated into the direct process. The resulting resonant Green's function is simply
\begin{equation}\label{res}
    G^{\rm (res)}({\bf r'},{\bf r'})=\frac{E^{\scs (0)}_x({\bf r}, k_y)E^{\scs (0)}_x({\bf r'}, -k_y)}
{2k[k-k(k_x)]},
\end{equation}
where $E^{\scs (0)}_x({\bf r}, k_y)$ is the profile of the resonant eigenmode. Above we omitted the band index of the dispersion $k(k_x)$ bearing in mind that the resonant approximation
uses the dispersion and the mode profiles of the BIC host band.

At first let us again consider linear scattering problem.
As before we assume that
a TE polarized plane wave with intensity $I_0$  impinges onto the structure at the near normal incidence. Then, solving Eq. (\ref{Helmholtz}) with the resonant Green's function Eq. (\ref{res}) one finds
\begin{equation}\label{res_solution}
    E_x=\frac{\sqrt{I_0}E^{\scs (0)}_x({\bf r}, k_y)}
{2k[k-k(k_y)]}\int_VdVE_x^{\scs (0)}({\bf r'}, -k_y)J({\bf r'}),
\end{equation}
where the souse term can be express through the incident field $E_x^{\rm (in)}$ as
\begin{equation}\label{J}
   J=- \left(\frac{\partial^2}{\partial y^2}+ \frac{\partial^2}{\partial z^2}\right)E_x^{\rm (in)}-n_0^2E_x^{\rm (in)},
\end{equation}
where $E_x^{\rm (in)}$ is normalized to carry a unit of energy per unit of time per unit are of the boundary of the scattering domain. Thus,
amplitude of the physical incident wave is only controlled by $I_0$.

To be consistent with the resonant approximation we also assume that
the near-field response is dominated by the quasi-BIC eigenmode
\begin{equation}\label{resonant_substitution}
E_x=\frac{1}{\sqrt{A}}aE_x^{\scs(0)},
\end{equation}
where $A$ is the normalization constant.
Substituting the above equation into Eq. (\ref{res_solution}) one finds
\begin{equation}\label{amplitude2}
a=\frac{\sqrt{I_0A}}
{2k[k-k(k_x)]}\int_VdVE_x^{\scs(0)}({\bf r'}, -k_y)J({\bf r'}).
\end{equation}
By comparing Eqs. (\ref{amplitude}) and (\ref{amplitude2}) one finds
\begin{equation}
    d_1=\frac{i\sqrt{A}c}
{2k}\int_VdVE_x^{\scs(0)}({\bf r'}, -k_y)J({\bf r'}),
\end{equation}
where $c$ is the speed of light.
The above equation is difficult to be applied in computations, since it
requires the explicit analytic form of the leaky mode profiles under normalization condition
Eq.~(\ref{norm}). We, however, already know from the previous subsection how $d_1$
can be found from the scattering spectra with the use of Eqs.~(\ref{d}) and (\ref{ab}).

\subsection{Nonlinear case}
Now let us generalize the above result onto the non-linear case. After taking the same route we end up with the following equation for $a$
\begin{equation}\label{amplitude3}
a+{\cal J}\frac{n_0n_2k}{A[k-k(k_y)]}|a|^2a=\frac{\sqrt{I_0}d_1}
{ic[k-k(k_y)]},
\end{equation}
where
\begin{equation}\label{J2}
    {\cal J}=\int_{V_{\rm nlin}}dVE_x^{\scs(0)}(-k_y)E_x^{\scs(0)}(k_y)\left|E_x^{\scs(0)}(k_y)\right|^2,
\end{equation}
and the integration is performed only over the domain with nonlinear refractive index
$V_{\rm nlin}$. Finally notice that $\cal{J}$ must be parabolic in $k_y$, since
the problem is symmetric with respect to the angle of incidence. Therefore we drop the dependence on $k_y$ using the field profile in the $\Gamma$-point, i.e.
the BIC
\begin{equation}\label{J_final}
    {\cal J}=\int_{V_{\rm nlin}}dV\left|E_x^{\rm \scs (BIC)}\right|^4 + {\cal O}(k_y^2).
\end{equation}
Since the BIC profile is a well behaved function decaying with
the distance from the grating. Following \cite{pankin2020fano} one easily finds that Eq. (\ref{norm}) is equivalent to the integration over $z$ from plus to minus infinity
\begin{equation}\label{norm_final}
  1=\int^{\infty}_{-\infty}dVn_0^2\left|E_x^{\rm \scs (BIC)}\right|^2.
\end{equation}
Importantly, Eq. (\ref{norm_final}) does not contain surface terms and, thus,
can be easily implemented in simulations taking into account that the BIC decays exponentially with $|z| \rightarrow  \infty $. Finally, notice that
the integral in Eq. (\ref{norm_final}) is equal to twice electromagnetic energy stored in the BIC. Thus, to be consistent with Eq. (\ref{cond3})
we set $A=2$.

\subsection{Nonlinear temporal coupled mode equation}
Now, in accordance with Eq. (\ref{amplitude3}) the final result reads
\begin{equation}\label{nlin_freq}
  [i(\omega-\omega_0)+\Gamma]a+i\frac{{\cal J}}{2}{n_0n_2\omega}|a|^2a={\sqrt{I_0}d_1}.
\end{equation}
Equation (\ref{nlin_freq}) only differs from Eq. (\ref{amplitude})
by the presence of nonlinear term proportional to $J$. It means that
for describing the nonlinear response in the spectral vicinity of a BIC
it is sufficiently to know the solution of the linear problem including
the field profile of the BIC. Once the BIC field profile is known it
can be substituted into Eq. (\ref{J_final}) to find $\cal J$. The nonlinear Eq. (\ref{nlin_freq}) can be then solved for the response
in the frequency domain.

It is remarkable that the nonlinear correction in Eq. (\ref{nlin_freq})
exactly coincides with that obtained previously with the perturbation
theory \cite{Bravo-Abad07}. Notice, however, that the results reported in
 \cite{Bravo-Abad07} have been obtained under assumption of
 smallness of the nonlinear term. Another issue with straightforward application of the perturbation theory is the normalization condition
 of the unperturbed eigenmodes. The formal solution presented in
 \cite{Bravo-Abad07} involves integration across the whole space
 which is impossible due to divergence of the resonant eigenmodes
 in the far zone \cite{Lalanne18}. As one can see from the previous subsection the normalization issue can only be easily resolved
 in the spectral vicinity of a BIC.

Finally, in the time domain Eq. (\ref{nlin_freq})
can be replaced by
\begin{equation}
      \frac{d}{dt}\left(a+\frac{{\cal J}}{2}{n_0n_2}|a|^2a\right)=(i\omega_0-\Gamma)a +{\sqrt{I_0}d_1}e^{i\omega t}.
\end{equation}
The time-harmonic solutions of the above equations can be tested
for stability by series expansions with respect to small perturbation
as explained in  \cite{Bulgakov19} .

\section{Numerical validation}

In this section we apply our previous findings to the scattering spectra. To obtain the matrices of the direct process we
numerically solved the linear scattering problem at exact normal
incidence with the vacuum wave number of the incident
wave equal to that of the BIC. For BIC 1 the following matrix of the direct process has been found
\begin{equation}\label{C1}
    \hat{C}_{\rm \scs BIC 1}=
    \left(\begin{array}{cc}
        -0.3753 + 0.0494i & -0.8834 + 0.2765i \\
         -0.8833 + 0.2765i & 0.3365 - 0.1734i
    \end{array}\right),
\end{equation}
while for BIC 2 the numbers are
\begin{equation}\label{C2}
    \hat{C}_{\rm \scs BIC 2}=
    \left(\begin{array}{cc}
       -0.9333 - 0.1121i &  0.0768 - 0.3323i \\
         0.0768 - 0.3323i & -0.8879 - 0.3086i
    \end{array}\right).
\end{equation}
In the next step the numerical values of the three remaining
parameters $\omega_0$, $\Gamma$, and $\alpha$ were evaluated by fitting
to the scattering spectra in Fig. \ref{fig2} at different angles of incidence. The effective nonlinearity
coefficient was found with Eq. (\ref{J_final}) by integrating the BIC profiles shown in Fig.~\ref{fig1}~(b), and  Fig.~\ref{fig1}~(c). The results are collected in
Table 1. In Table 1 we also present
the ratio of the absolute values of the coupling constants $d_1$ and $d_2$. One can see that in both cases the coupling to the lower half-space
is somewhat larger than to the upper half-space, $|d_1|<|d_2|$.

\begin{table}[h]
{\bf \caption{List of the TCMT parameters.}}\begin{center}
\begin{tabular}{p{0.5in}ccccccc}
\hline
\ & $\theta$  & $h\omega_0/c$ & $h\Gamma/c$ & $\alpha$ (p.d.u.)
 & $|d_1/d_2|$ & ${\cal J}$ (p.d.u.) &
\\ \hline
BIC 1 & $1^{\rm o}$& $2.5670$ & $0.065\cdot 10^{-5}$ & $6.24\cdot 10^{-3}$
  & $0.907$ & $3.04\cdot 10^{-2}$ \\
BIC 1 & $2^{\rm o}$ & $2.5689$ & $0.031\cdot 10^{-3}$ & $1.23\cdot 10^{-2}$
&  $0.914$ &  $3.04\cdot 10^{-2}$\\
BIC 2 & $0.2^{\rm o}$  & $4.39074$ & $1.59\cdot 10^{-3}$ & $9.05\cdot 10^{-2}$
& $0.717$ & $3.79 \cdot 10^{-4}$
  \\ \hline
\end{tabular}
\end{center}
\end{table}

\begin{figure}[t]
\centering\includegraphics[width=1\textwidth, height=0.4\textwidth, trim={4cm 1.5cm 3cm 1.5cm},clip]{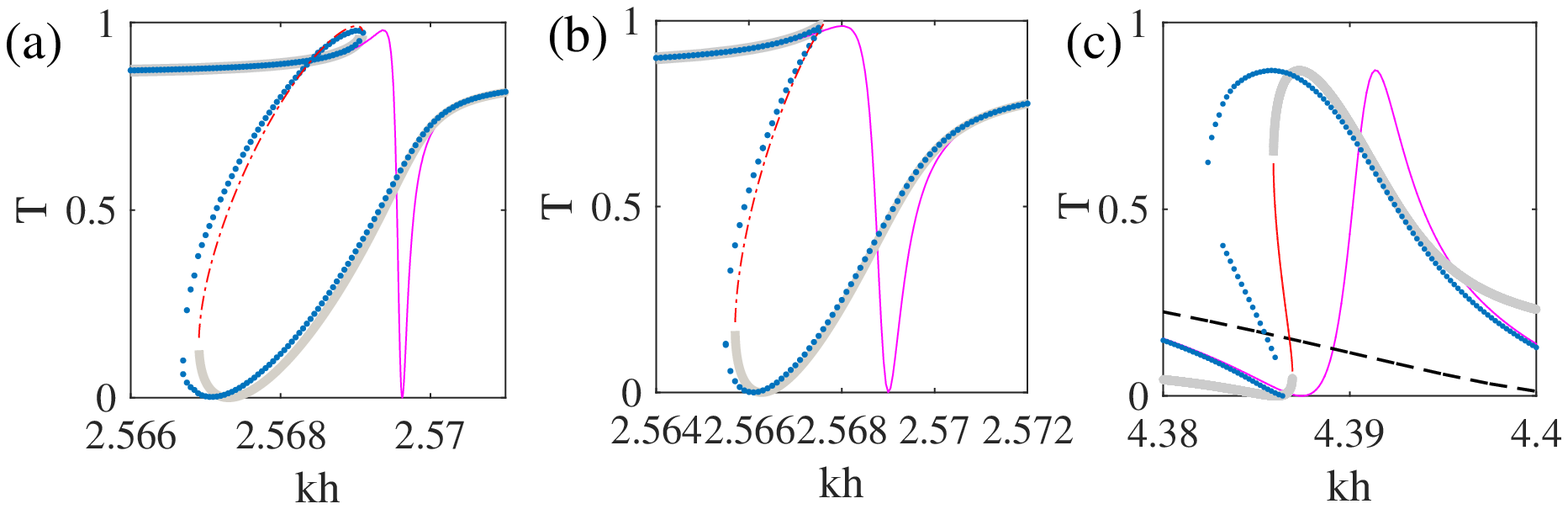}
\caption{Bistability in transmittance spectra. (a) BIC 1, $\theta=1^{\rm o}$, $I_0=1.82\cdot 10^{5}~{\rm MW/m^2}$. (b) BIC 1, $\theta=2^{\rm o}$, $I_0=8.39\cdot 10^{5}~{\rm MW/m^2} $. (c) BIC 2, $\theta=0.2^{\rm o}$, $I_0=3.92\cdot 10^{8}~{\rm MW/m^2}$. The thin magenta lines demonstrate Fano resonances unperturbed by the nonlinearity. The blue dots are full-wave numerical solutions obtained with pseudospectral method. The stable solutions of Eq. (\ref{nlin_freq}) are shown by thick grey lines. Thin red lines show unstable solutions of Eq. (\ref{nlin_freq}). Dashed black line in (c)
show the transmittance at exact normal incidence.}\label{fig3}
\end{figure}

To obtain the nonlinear scattering spectra Eq. (\ref{Helmholtz}) was solved numerically with the pseudospectral method \cite{Yuan13}. In our simulations we took $n_2=5\cdot 10^{-18}~\rm{m^2/W}$ which corresponds to silicon at $1.8~{\rm \mu m}$ \cite{Yue12}. The results are plotted
in Fig. \ref{fig3} in comparison with numerical solution of Eq. (\ref{nlin_freq}). The intensities of the incident waves are
given in the caption to Fig. \ref{fig3}. First of all one can see in Fig.~\ref{fig3}~(a) and Fig.~\ref{fig3}~(b)
that there is a reasonably good agreement between the TCM and the full-wave spectra for BIC 1. The agreement can be made perfect
by slightly tuning $\alpha$ and/or $\cal{J}$. For BIC 2, however, the agreement is not that good and our theory can only provide a rough estimate of the
parameter values leading to optical bistability. The discrepancy
is due to the single-mode approximation which is not capable
to account for all features of the BIC emerging with an avoided crossing. To
highlight the limitations of the single-mode TCMT in Fig. \ref{fig3} (c)
we plotted the transmittance at the normal incidence. One
can easily see that the transmittance at the normal incidence is dependent on frequency, whereas the single-mode TCMT assumes that it is constant. On the other hand, even the single-mode approximation
manages to grasp the major feature of BIC 2 with respect to initiating
bistability. One can see from Table 1 that the effective nonlinearity
coefficient $\cal{J}$ is two orders of magnitude smaller with BIC 2 than
with BIC 1. The reason for this is clearly seen from Fig.~\ref{fig1}~(b) and Fig.~\ref{fig1}~(c) - the field of BIC 1 is concentrated about the nonlinear
medium (silicon bars), while for BIC 2 the field is evenly spread
across the whole grating. The small value of  $\cal{J}$ results in very
high intensities needed to trigger optical bistability. This rules out
application of BIC 2 in a realistic experiment.

\section{Summary and conclusions}
In this paper we considered the effect of optical bistability induced
by bound states in the continuum (BICs) in dielectric gratings. We proposed a coupled mode approach which leads to
a single nonlinear equation for the amplitude of the resonant eigenmode
of the BICs host band. It is shown how all parameters entering the nonlinear coupled mode equation can be evaluated from the solution
of the linear scattering problem.

We believe that the approach presented here can be of use in engineering
photonic systems with the resonantly enhanced nonlinear response, as  the
coupled mode equation is much easier to get solved than nonlinear Maxwell's equations. At the same time our approach gives a cue for
choosing the type of BICs in order to maximize the nonlinear effect. Namely, it has been shown that the BIC with the frequency lower than
the first diffraction order in the substrate are better in activating
the nonlinearity.

On the other hand, from the fundamental view point we have seen
that the scattering of light in the spectral vicinity of a BIC can
be handled by the resonant state expansion method \cite{Weiss17}.
This naturally prompts us to extend the theory for the two-mode
case which potentially leads to an intricate interplay between
the BICs and the other mode with a finite live-time that
can be excited from the far zone even at the normal incidence \cite{Krasikov18}. The application of resonant state expansion
to finite-lived states would, however, require introducing the
analytical continuation normalization condition, Eq. (\ref{continuation}) to eigenmodes that are known only numerically. We speculate that the above
problem can be an interesting topic for future studies.

\section*{Acknowledgments}
 This work was financially supported by the Government of the Russian Federation through
 the ITMO Fellowship and Professorship Program.

\bibliography{BSC_light_trapping}


\end{document}